\documentstyle[12pt,a41,axodraw]{article}

\newcommand\beq{\begin{equation}}
\newcommand\eeq{\end{equation}}
\newcommand\bea{\begin{eqnarray}}
\newcommand\eea{\end{eqnarray}}

\begin{document}
\thispagestyle{empty}

\mbox{}
\begin{flushleft}
DESY  98-006 \hfill {\tt hep-ph/9801325}\\
January 1998\\
\end{flushleft}
\vspace*{\fill}
\begin{center}
{\LARGE\bf Second order QCD contributions to}

\vspace{2mm}
{\LARGE\bf  polarized spacelike and timelike}

\vspace{2mm}
{\LARGE\bf processes}

\vspace*{20mm}
\large
{W.L. van Neerven \footnote{Talk presented at the "Cracow Epiphany
Conference on Spin Effects in Particle Physics and Tempus Workshop"
Cracow, Poland, January 9-11, 1998.}}
\\

\vspace{2em}

\normalsize
%{\it DESY--Zeuthen, Platanenallee 6,}\\
%{\it D--15735 Zeuthen,  Germany}
{\it DESY-Zeuthen, Platanenallee 6, D-15738 Zeuthen, Germany}%
\footnote{
On leave of absence from Instituut-Lorentz, University of Leiden,
P.O. Box 9506, 2300 RA Leiden, The Netherlands.}

\vspace{4cm}
%\today
\end{center}
\vspace*{\fill}
\begin{abstract}
\noindent
We will give an outline of the computation of the QCD corrections to the
spin structure function $g_1(x,Q^2)$ and the spin fragmentation function
$g_1^{\rm H}(x,Q^2)$ which are measured in deep inelastic electron-proton
scattering and in electron-positron annihilation respectively. In
particular
we show how to deal with the $\gamma_5$-matrix and the Levi-Civita tensor,
appearing in the amplitudes of the parton subprocesses, when the method of
$N$-dimensional regularization is used.
\end{abstract}
\vspace*{1cm}

\vspace*{\fill}
%\end{document}

\section{Deep inelastic electron-proton scattering}
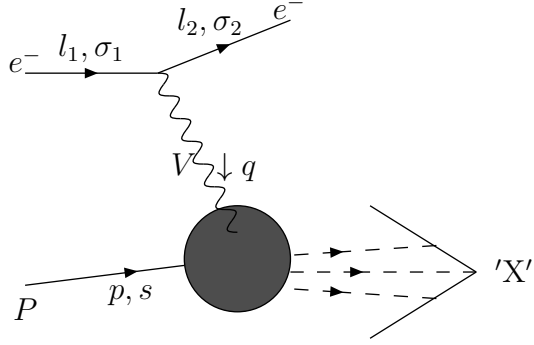
\begin{figure}
\begin{center}
  \begin{picture}(185,130)(0,0)
    \ArrowLine(0,20)(80,30)
    \DashArrowLine(80,30)(155,35){5}
    \DashArrowLine(80,25)(170,25){5}
    \DashArrowLine(80,20)(155,15){5}
    \Line(130,50)(170,25)
    \Line(130,0)(170,25)
    \GCirc(80,30){20}{0.3}
    \Photon(50,100)(80,40){3}{7}
    \ArrowLine(0,100)(50,100)
    \ArrowLine(50,100)(100,120)
    \Text(0,110)[t]{$e^-$}
    \Text(100,130)[t]{$e^-$}
    \Text(0,15)[t]{$P$}
    \Text(60,70)[t]{$V$}
    \Text(185,30)[t]{$'{\rm X}'$}
    \Text(40,20)[t]{$p,s$}
    \Text(25,115)[t]{$l_1,\sigma_1$}
    \Text(70,125)[t]{$l_2,\sigma_2$}
    \Text(80,70)[t]{$\downarrow q$}
  \end{picture}
  \caption[]{Kinematics of polarized deep inelastic
             electron-proton scattering}
  \label{fig:1}
\end{center}
\end{figure}

Deep inelastic electron-proton scattering proceeds via the following 
reaction (see Fig. \ref{fig:1})
\begin{eqnarray}
\label{eqn:1}
e^-(l_1,\sigma_1) + P(p,s) \rightarrow e^-(l_2,\sigma_2) + '{\rm X}'\,.
\end{eqnarray}
Here $'{\rm X}'$ denotes any inclusive hadronic final state and $V$ 
in Fig. \ref{fig:1} stands
for the neutral intermediate vector bosons given by $\gamma,Z$. For 
simplicity we 
will assume that the momentum transfer is very small with respect to the mass
of the Z-boson so that the process in Fig. \ref{fig:1}
is dominated by the one photon exchange mechanism only.
In the case the proton is polarized
parallel ($\rightarrow$) or anti-parallel ($\leftarrow$) with respect to the 
spin of the incoming electron we obtain the cross section
\begin{eqnarray}
\label{eqn:2}
\frac {d^2\sigma(\rightarrow)}{dx\,dy}-\frac {d^2\sigma(\leftarrow)}{dx\,dy}
=\frac{4 \pi \alpha^2}{Q^2}\Big[ \{ 2 - y \} g_1(x,Q^2) \Big]\,, 
\end{eqnarray}
where $g_1(x,Q^2)$ denotes the longitudinal
spin structure function. Further we have defined the scaling variables
\begin{eqnarray}
\label{eqn:3}
x=\frac{Q^2}{2p \cdot q} \qquad , \qquad 
y = \frac{p \cdot q}{p \cdot l_1} \qquad , \qquad q^2=-Q^2 < 0
\,.
\end{eqnarray}
%fig1
The spin structure functions show up in the antisymmetric part of the hadronic
tensor
\begin{eqnarray}
\label{eqn:4}
W_{\mu\nu}(p,q,s) = \frac{1}{4\pi} \int\, d^4 z \,e^{iq\cdot z}
\langle p,s \mid J_{\mu}(z) J_{\nu}(0) \mid p,s \rangle \,,
\end{eqnarray}
which is given by
\begin{eqnarray}
\label{eqn:5}
W_{\mu\nu}^A (p,q,s)=\frac{m}{2p\cdot q} \epsilon_{\mu\nu\alpha\beta} 
q^{\alpha}
\Big [ s^{\beta} g_1(x,Q^2) + ( s^{\beta}- \frac{s\cdot q}{p\cdot q} 
p^{\beta})
 g_2(x,Q^2) \Big ] \,.
\end{eqnarray}
Here $g_2(x,Q^2)$ denotes the transverse spin structure function which is
kinematically suppressed in cross section (\ref{eqn:2}).
Since the leading power corrections are of twist two, one can give a parton 
model description of the longitudinal structure function which
can be written as
\begin{eqnarray}
\label{eqn:6}
&& g_1(x,Q^2) =
 \frac{1}{n_f} \sum_{k=1}^{n_f} e_k^2 \int_x^1 \frac{dz}{z}
\Big [ \Delta f_q^{\rm S}(\frac{x}{z},\mu^2)  
\Delta {\cal C}_{1,q}^{\rm S}(z,\frac{Q^2}{\mu^2})
\nonumber\\[2ex]
&& + \Delta f_g^{\rm S}(\frac{x}{z},\mu^2)  
\Delta {\cal C}_{1,g}^{\rm S}(z,\frac{Q^2}{\mu^2})
+ n_f \Delta f_{q,k}^{\rm NS}(\frac{x}{z},\mu^2) 
\Delta {\cal C}_{1,q}^{\rm NS}
(z,\frac{Q^2}{\mu^2}) \Big ]\,.
\end{eqnarray}
Unfortunately there does not exist such a simple formula for $g_2(x,Q^2)$ 
because of twist three contibutions which are not power suppressed with
respect to the twist two parts. Hence one cannot give a simple parton model
interpretation for the transverse spin structure function and we will  
therefore not discuss it in the subsequent part of this paper.\\
In Eq. (\ref{eqn:6}) we have used the following notation. The charge of
the light quarks is denoted by $e_k$ and $n_f$ stands for the number of light 
flavours. The spin parton densities $\Delta f_i(z,\mu^2)$ ($i=q,g$) depend on
factorization scale $\mu$ which is put to be equal to the renormalization 
scale. The spin parton coefficient functions $\Delta {\cal C}_{1,i}$
depend on the same scale $\mu$. The quark parton densities and the quark
coefficient functions can be split in non-singlet (${\rm NS}$) and 
singlet (${\rm S}$) parts with respect to the flavour group. The singlet and 
non-singlet combinations of parton densities are given by
\begin{eqnarray}
\label{eqn:7}
\Delta f_q^{\rm S}(z,\mu^2) & = & \sum_{k=1}^{n_f} [ \Delta f_k(z,\mu^2)+
\Delta f_{\bar k}(z,\mu^2)]
\,,
\end{eqnarray}
and
\begin{eqnarray}
\label{eqn:8}
\Delta f_{q,k}^{\rm NS} (z,\mu^2) & = & \Delta f_k(z,\mu^2) + 
\Delta f_{\bar k}(z,\mu^2)
-\frac{1}{n_f} \Delta f_q^{\rm S}(z,\mu^2) \,,
\end{eqnarray}
respectively. 
Since it turns out that the 
equations are easier to study when one performs a 
Mellin transform defined by
\begin{eqnarray}
\label{eqn:9}
F^{(n)} = \int_0^1 dz \, z^{n-1} F(z) \,,
\end{eqnarray}
we will present all the following formulae in this representation.\\
The parton densities and the coefficient functions above satisfy the 
renormalization group equations. Let us first define the differential operator
\begin{eqnarray}
\label{eqn:10}
D\,\raisebox{-3mm}{$\stackrel{\displaystyle =}{\scriptstyle D}$} \,
\mu \frac{\partial}{\partial \mu} + \beta(g) \frac{\partial}{\partial g}
\qquad , \qquad \beta(g) = - \beta_0 \frac{g^3}{16 \pi^2} + \cdots \,.
\end{eqnarray}
Using this notation the renormalization group equations for the parton 
densities read
\begin{eqnarray}
\label{eqn:11}
D\, \Delta f_{q,k}^{\rm NS,(n)}&=&- \Delta \gamma_{qq}^{\rm NS,(n)} 
\Delta f_{q,k}^{\rm NS,(n)} \qquad , \qquad k=u,d \cdots
\nonumber\\[2ex]
D\, \Delta f_{i}^{\rm S,(n)}&=&- \Delta \gamma_{ij}^{\rm S,(n)} 
\Delta f_{j}^{\rm S,(n)} \qquad , \qquad i,j=q,g \quad,
\end{eqnarray}
and for the coefficient functions 
\begin{eqnarray}
\label{eqn:12}
D\, \Delta {\cal C}_{1,q}^{\rm NS,(n)}&=& \Delta \gamma_{qq}^{\rm NS,(n)} 
\Delta {\cal C}_{1,q}^{\rm NS,(n)}
\nonumber\\[2ex]
D\, \Delta {\cal C}_{1,i}^{\rm S,(n)}&=& \Delta \gamma_{ji}^{\rm S,(n)}    
\Delta {\cal C}_{1,j}^{\rm S,(n)} \qquad , \qquad i,j=q,g \quad.
\end{eqnarray}
From the equations above it follows that the structure function is a 
renormalization group invariant i.e. 
\begin{eqnarray}
\label{eqn:13}
D~g_1^{(n)}(Q^2)=0 \,, 
\end{eqnarray}
which implies that it is a physical quantity independent of the scale $\mu$.

The anomalous dimensions and the coefficient functions are calculable order
by order in perturbation theory. Let us first sketch the derivation of the
anomalous dimensions before we pay attention to the coefficient functions.
In \cite{mene} the anomalous dimensions appearing in the spin dependent 
quantities have been derived from the calculation of the
operator matrix elements (OME's). For an alternative derivation see 
\cite{vogel}. These OME's are obtained by sandwiching local operators 
between quark and gluon states.
These operators appear in the lightcone expansion of the product of the 
electromagnetic currents in Eq. (\ref{eqn:4}).
Suppressing some irrelevant Lorentz indices the expansion reads as follows
\begin{eqnarray}
\label{eqn:14}
J(x)J(0)
\raisebox{-3mm}{$\stackrel{\displaystyle =}{\scriptstyle x^2 \rightarrow 0}$} 
\sum_{n=0}^{\infty} \sum_i c_{1,i}^{(n)}(x^2) O_i^{(n)}(0) \qquad , \qquad 
i=q,g \quad.
\end{eqnarray}
Here $n$ denotes the spins of the local operators $O_i^{(n)}$ and 
$c_{1,i}^{(n)}(x^2)$ 
are the Fourier transforms of the coefficient functions in position space. 
The operators of twist two, which can also be split into singlet
and non-singlet parts, are given in the literature 
(see e.g. Eqs. (2.5)-(2.7) in \cite{mene}).
Since, in the Bj{\o}rken limit, the integrand in Eq. (\ref{eqn:4}) is 
dominated by the light
cone we can replace the current-current product by the above expansion so
that one has to compute the renormalized OME's
\begin{eqnarray}
\label{eqn:15}
A_{ij}^{r,(n)}(\frac{-p^2}{\mu^2}) 
= \langle j(p) \mid O_i^{(n)} \mid p(j) \rangle \,,
\end{eqnarray}
where $i,j=q,g$ and $p$ denotes the external momentum of the quarks and gluons
and $r={\rm NS,S}$.
In \cite{mene} and recently also in \cite{msn} the above 
operator matrix element (OME) has been computed up to second order in the 
strong coupling 
constant $\alpha_s$. The calculation proceeds as follows. After having 
derived 
the operator vertices (see Appendix A in \cite{mene}) one has to compute
the Feynman graphs (see Figs. 1-6 in \cite{mene}) which correspond to the
unrenormalized (bare) OME's. The latter 
reveal ultraviolet singularities which are regularized by 
N-dimensional regularization. The unrenormalized OME's indicated by a hat
can be written in the form
\begin{eqnarray}
\label{eqn:16}
\hat A_{ij}^{r,(n)}(\frac{-p^2}{\mu^2},\frac{1}{\varepsilon})&=&  \delta_{ij} 
+ \hat a_s S_{\varepsilon}
\big(\frac{-p^2}{\mu^2}\big)
^{\varepsilon/2} \Big [ \frac{1}{\varepsilon} \Delta \gamma_{ij}^{(n),(0)} 
+ \Delta a_{ij}^{(n),(1)}
+ \varepsilon \Delta a_{ij}^{\varepsilon,(n),(1)}\Big ]
\nonumber\\[2ex]
&&+{\hat a}_s^2 S_{\varepsilon}^2
\big(\frac{-p^2}{\mu^2}\big)^{\varepsilon} \Big[
 \frac{1}{\varepsilon^2} \Big \{ \frac{1}{2} \Delta \gamma_{ik}^{(n),(0)} 
 \Delta \gamma_{kj}^{(n),(0)}
- \beta_0 \Delta \gamma_{ij}^{(n),(0)} \Big \}
\nonumber\\[2ex]
&& +\frac{1}{\varepsilon} \Big \{ \frac{1}{2} \Delta \gamma_{ij}^{(n),(1)}
 - 2 \beta_0 \Delta a_{ij}^{(n),(1)} + \Delta \gamma_{ik}^{(n),(0)} \Delta 
a_{kj}^{(n),(1)} \Big \}
\nonumber\\[2ex]
&& +\Delta  a_{ij}^{(n),(2)} - 2 \beta_0 \Delta a_{ij}^{\varepsilon,(n),(1)}
+ \Delta \gamma_{ik}^{(n),(0)} \Delta a_{kj}^{\varepsilon,(n),(1)} \Big ]\,,
\end{eqnarray}
where $S_{\varepsilon}$ is the spherical factor characteristic of 
$N$-dimensional regularization.
Here the hat indicates that all quantities are unrenormalized with respect to
coupling
constant and operator renormalization. The algebraic structure shown by the 
expression above follows from the renormalization group. In addition to the
anomalous dimensions one also encounters the coefficients of the 
beta-function.
For instance $\beta_0$ (see Eq. (\ref{eqn:10})) is the lowest order 
coefficient, which also appears in the coupling constant renormalization,
given by
\begin{eqnarray}
\label{eqn:17}
\hat a_s = a_s(\mu^2) \Big [ 1 + a_s(\mu^2) S_{\varepsilon} \Big \{ 2 \beta_0
\frac{1}{\varepsilon} \Big \} \Big ] \qquad,\qquad  a_s=\frac{\alpha_s}{4\pi}\,.
\end{eqnarray}
From the expression above one can in principle extract the first and second 
order anomalous dimension of the local operators in Eq. (\ref{eqn:14}) 
which are given by
$\Delta \gamma_{ij}^{(n),(0)}$ and $\Delta \gamma_{ij}^{(n),(1)}$
respectively. However in the renormalization of the
OME's one has to deal with two difficulties. The first one is caused by the 
fact
that usually the external momentum $p$ is taken off-shell ($p^2 < 0$). This
means that the OME in Eq. (\ref{eqn:15}) ceases to be a genuine S-matrix 
element and it becomes
gauge dependent. Therefore one also has to carry out gauge parameter 
renormalization. The second problem, which is characteristic of spin 
operators,
is the appearance of the $\gamma_5$-matrix and the Levi-Civita tensor
$\epsilon_{\mu\nu\alpha\beta}$ in the operator vertices (see Appendix A in 
\cite{mene}). In the case of $N$-dimensional regularization one has to find a 
suitable prescription to define these essentially four dimensional  
quantities in $N$-dimensions. In \cite{mene} and \cite{msn} the vertex 
$\gamma_{\mu}\gamma_5$ is replaced by
\begin{eqnarray}
\label{eqn:18}
\gamma_{\mu} \gamma_5 = \frac{i}{6} \epsilon_{\mu \alpha\beta\sigma}
\gamma^{\alpha}\gamma^{\beta}\gamma^{\sigma} \,,
\end{eqnarray}
so that only the Levi-Civita tensor appears in the OME's. This prescription
is equivalent to the one given by 't Hooft and Veltman \cite{hove} which is 
worked out in more detail by Breitenlohner and Maison \cite{brma} (HVBM).
For the replacement of the $\gamma_5$-matrix in Eq. (\ref{eqn:18}) see
\cite{akde}, \cite{larin}. Although this prescription preserves the cyclicity
of the traces it destroys the anticommutativity of the $\gamma_5$-matrix.
This will mean that some Ward-identies or theorems will be violated.
For example the non-singlet axial vector current, presented by the operator
$O_q^{{\rm NS},(1)}$ in Eq. (\ref{eqn:14}), gets renormalized in second order 
although it is conserved. Furthermore the Adler-Bardeen theorem \cite{adba} 
concerning
the nonrenormalization of the Adler-anomaly is violated. This will affect 
the renormalization of the singlet axial-vector operator $O_q^{{\rm S},(1)}$ 
in order $\alpha_s^3$. 
To undo these effects
one has to introduce an additional renormalization constant in order to obtain
the correct anomalous dimensions in the ${\overline {\rm MS}}$-scheme. 
The latter have now to be extracted from the
renormalized rather than the unrenormalized OME's. After coupling constant 
renormalization the OME's are renormalized as follows
\begin{eqnarray}
\label{eqn:19}
{\bar A}_{qq}^{{\rm NS},(n)}&=&{\bar Z}_{qq}^{5,{\rm NS},(n)}
({\bar Z}^{-1})_{qq}^{{\rm NS},(n)}
{\hat A}_{qq}^{{\rm NS},(n)}
\nonumber\\[2ex]
{\bar A}_{ij}^{{\rm S},(n)}&=&Z_{qq}^{5,{\rm S},(n)} (Z^{-1})_{iq}^{{\rm S},(n)}
{\hat A}_{qj}^{{\rm S},(n)}+({\bar Z}^{-1})_{ig}^{{\rm S},(n)} 
{\hat A}_{gj}^{{\rm S},(n)} \,,
\end{eqnarray}
where we have chosen the ${\overline {\rm MS}}$-scheme. In this scheme the 
constant for the HVBM-prescription can be written up to order $\alpha_s^2$ as
\begin{eqnarray}
\label{eqn:20}
Z_{qq}^{5,r,(n)}(\frac{1}{\varepsilon}) = 1 
+ a_s S_{\varepsilon} \Big [ z_{qq}^{(n),(1)} \Big ]
+ a_s^2 S_{\varepsilon}^2 \Big [  \frac{1}{\varepsilon} \beta_0
z_{qq}^{(n),(1)} + z_{qq}^{r,(n),(2)}  \Big ] \,,
\end{eqnarray}
with $r={\rm NS,S}$. Notice that the difference between the singlet (S) and 
the non-singlet (NS) expression for $Z_{qq}^{5,r}$ shows up for the first 
time in second order (see \cite{larin}). In this reference $Z_{qq}^{5,r}$ 
has been calculated for the first moment ($n=1$) up to order $\alpha_s^3$.
Recently this constant has been computed up to second order in \cite{msn} but 
now for general moments. In the non-singlet case it can be computed from the
ratio 
\begin{eqnarray}
\label{eqn:21}
Z_{qq}^{5,{\rm NS},(n)}(\frac{1}{\varepsilon}) = 
\frac{\hat A_{qq}^{{\rm NS},(n)}(-p^2/\mu^2,1/\varepsilon)\mid_{\rm naive}}
{\hat A_{qq}^{{\rm NS},(n)}(-p^2/\mu^2,1/\varepsilon)\mid_{\rm HVBM}}
\mid_{p^2=-\mu^2} \,,
\end{eqnarray}
where in the numerator one has used the so called naive prescription in which
the $\gamma_5$-matrix anticommutes with all other $\gamma$-matrices 
irrespective of the value for the dimension $N$. The use of the naive method 
implies that the numerator can be replaced by the
spin averaged OME $\hat A_{qq}^{{\rm NS},(n)}$ in which the $\gamma_5$-matrix
does not appear. A similar derivation exists for 
$Z_{qq}^{5,{\rm S},(n)}$ where one also makes a comparison between the
naive $\gamma_5$ and the HVBM prescription. From the considerations presented
above one could have asked the question why it is preferable to choose the
HVBM instead of the naive prescription since in the latter case 
$Z_{qq}^{5,r,(n)}=1$? The reason is that 
the Levi-Civita tensor appears in the OME ${\hat A}_{gq}$ which induces in 
the subgraphs containing 
quark lines the HVBM prescription. Therefore the naive method is inconsistent
and it is better to use a consistent procedure like HVBM where all constants
are fixed once and for all.
The operator renormalization constants in Eq. (\ref{eqn:19}), presented in 
the ${\overline {\rm MS}}$-scheme, read as follows
\begin{eqnarray}
\label{eqn:22}
&& ({\bar Z}^{-1})_{ij}^{r,(n)}(\frac{1}{\varepsilon})=\delta_{ij}
+  a_s S_{\varepsilon}
\Big [ - \frac{1}{\varepsilon} \Delta \gamma_{ij}^{(n),(0)} \Big ]
\nonumber\\[2ex]
&& +  a_s^2 S_{\varepsilon}^2
\Big [  \frac{1}{\varepsilon^2} \Big \{\frac{1}{2} \Delta \gamma_{ik}^{(n),(0)}
\Delta \gamma_{kj}^{(n),(0)} - \beta_0 \Delta \gamma_{ij}^{(n),(0)} \Big \}
\nonumber\\[2ex]
&& + \frac{1}{2\varepsilon} \Big \{ \Delta \bar{\gamma}_{ij}^{(n),(1)} \pm
 \Delta \gamma_{ik}^{(n),(0)} \Delta z_{kj}^{(n),(1)} \Big \} \Big ] \,.
\end{eqnarray}
The above expression differs from the usual one by the appearance of the term
$z_{kj}^{(n),(1)}$ with $k=j=q$ which only contributes to
$({\bar Z}^{-1})_{qg}$ (plus sign) and $({\bar Z}^{-1})_{gq}$ (minus sign) up
to order $\alpha_s^2$. If this term is omitted 
then the anomalous dimensions will equal those present in 
Eq. (\ref{eqn:16}),
which differ by a finite renormalization from the ones presented in the 
${\overline {\rm MS}}$-scheme. Notice that the lowest order coefficients
$\Delta \gamma^{(n),(0)}$ are not affected by any $\gamma_5$ prescription. 
Finally we want to
emphasize that due to the pole term in Eq. (\ref{eqn:20}) $Z_{qq}^{5,r}$ 
does not represent a finite renormalization constant in the usual sense. 
Using the above procedure one can write 
the following expression for the renormalized OME in the 
${\overline {\rm MS}}$-scheme
\begin{eqnarray}
\label{eqn:23}
{\bar A}_{ij}(\frac{-p^2}{\mu^2}) &=& \delta_{ij}
+ a_s \Big [ \frac{1}{2} \Delta \gamma_{ij}^{(n),(0)}
\ln \Big ( \frac{-p^2}{\mu^2} \Big ) + \Delta {\bar a}_{ij}^{(n),(1)} \Big ]
\nonumber\\[2ex]
&&+a_s^2 \Big [ \Big \{ \frac{1}{8} \Delta \gamma_{ik}^{(n),(0)}
\Delta \gamma_{kj}^{(n),(0)}- \frac{1}{4} \beta_0 \Delta \gamma_{ij}^{(n),(0)}
\Big \} \ln^2 \Big ( \frac{-p^2}{\mu^2} \Big ) + \Big \{
\frac{1}{2} \Delta {\bar\gamma}_{ij}^{(n),(1)}
\nonumber\\[2ex]
&& -\beta_0 \Delta {\bar a}_{ij}^{(n),(1)}
+ \frac{1}{2}\Delta \gamma_{ik}^{(n),(0)} \Delta {\bar a}_{kj}^{(n),(1)} \Big \}
\ln \Big ( \frac{-p^2}{\mu^2})
+ \Delta {\bar a}_{ij}^{(n),(2)} \Big ] \,.
\end{eqnarray}
Notice that the coefficients $\Delta {\bar a}_{ij}^{(n),(k)}$ in this 
expression differ from the $\Delta a_{ij}^{(n),(k)}$ present 
in Eq. (\ref{eqn:16}). From the relation
\begin{eqnarray}
\label{eqn:24}
\Delta f_i^{(n)}(\mu^2) = {\bar A}_{ij}^{(n)}\Big (\frac{-p^2}{\mu^2}\big ) 
\Delta f_i^{(n)}(-p^2) \,,
\end{eqnarray}
and Eq. (\ref{eqn:11}) one concludes that the renormalized 
${\bar A}_{ij}^{(n)}$ satisfy  
the renormalization group equation given by
\begin{eqnarray}
\label{eqn:25}
D {\bar A}_{ij}^{(n)} = - \Delta {\bar \gamma}_{ik}^{(n)} {\bar A}_{kj}^{(n)}\,.
\end{eqnarray}
%---------------------------
%fig2
\begin{figure}
\begin{center}
  \begin{picture}(60,60)
  \Photon(30,0)(30,20){3}{3}
  \ArrowLine(0,20)(60,20)
  \end{picture}
\hspace*{1cm}
%  \hfill
  \begin{picture}(60,60)
  \GlueArc(30,10)(25,25,155){3}{7}
  \Photon(30,0)(30,20){3}{3}
  \ArrowLine(0,20)(60,20)
\end{picture}
\hspace*{1cm}
%  \hfill
  \begin{picture}(60,60)
  \Gluon(10,20)(40,40){3}{7}
  \Photon(30,0)(30,20){3}{3}
  \ArrowLine(0,20)(60,20)
\end{picture}
\hspace*{1cm}
%  \hfill
  \begin{picture}(60,60)
  \GlueArc(30,10)(25,20,160){3}{7}
  \GlueArc(30,10)(17,35,145){3}{4}
  \Photon(30,0)(30,20){3}{3}
  \ArrowLine(0,20)(60,20)
\end{picture}
\end{center}
\end{figure}
%--------------------------------------------
\begin{figure}
\begin{center}
  \begin{picture}(60,60)
  \GlueArc(30,10)(25,20,160){3}{7}
  \Gluon(15,20)(50,50){3}{7}
  \Photon(30,0)(30,20){3}{3}
  \ArrowLine(0,20)(60,20)
\end{picture}
\hspace*{1cm}
%  \hfill
  \begin{picture}(60,60)
  \Gluon(10,20)(50,50){3}{7}
  \Gluon(40,20)(60,35){3}{3}
  \Photon(30,0)(30,20){3}{3}
  \ArrowLine(0,20)(60,20)
\end{picture}
\hspace*{1cm}
%  \hfill
  \begin{picture}(60,60)
  \Gluon(10,20)(30,40){3}{7}
  \ArrowLine(55,50)(30,40)
  \ArrowLine(30,40)(55,30)
  \Photon(30,0)(30,20){3}{3}
  \ArrowLine(0,20)(60,20)
\end{picture}
\hspace*{1cm}
%  \hfill
  \begin{picture}(60,60)
  \Gluon(10,20)(30,40){3}{7}
  \ArrowLine(55,50)(30,40)
  \ArrowLine(30,40)(55,30)
  \Photon(45,0)(45,35){3}{3}
  \ArrowLine(0,20)(60,20)
\end{picture}
\vspace*{5mm}
  \caption{Contributions to the proces $\gamma^* + q \rightarrow
           '{\rm X}'$ contributing to the partonic structure function
           $\hat g_{1,q}$.}
  \label{fig:2}
\end{center}
\end{figure}
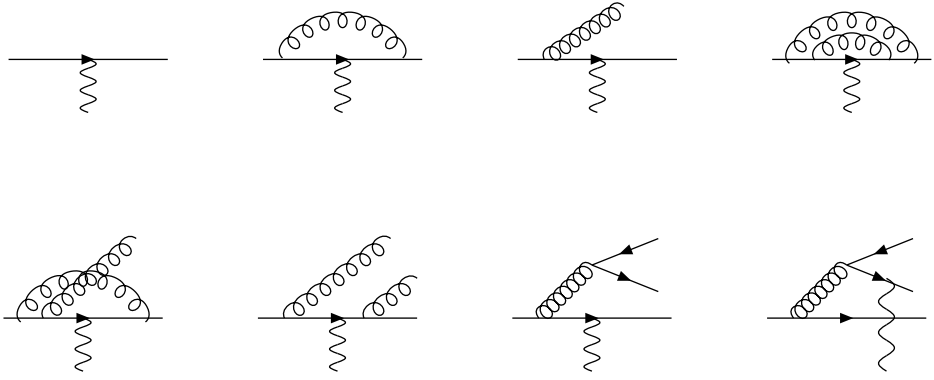
%---------------------------
After having discussed the renormalization of the OME's
we now explain the procedure to compute the coefficient functions
appearing in the spin structure function of Eq. (\ref{eqn:6}). 
They are obtained from the partonic
subprocesses denoted by
\begin{eqnarray}
\label{eqn:26}
\gamma^* + i \rightarrow '{\rm X}' \,,
\end{eqnarray}
where $i$ stands either for a quark (q) or a gluon (g) and $'{\rm X}'$
represents an inclusive multi-partonic state.
The above processes have been calculated up to order $\alpha_s^2$
in \cite{zine}. Some Feynman graphs are depicted in Fig. \ref{fig:2} ($i=q$) 
and in Fig. \ref{fig:3} ($i=g$).
The computation of the QCD corrections reveal ultraviolet, infrared and
collinear divergences which appear in the Feynman integrals and in the 
phase space
integrals. Like in the case of the operator matrix elements we regularize
them by means of $N$-dimensional regularization. Further we use the HVBM
prescription for the $\gamma_5$-matrix as discussed below Eq. (\ref{eqn:18}).
Adding all contributions one observes that the infrared singularities cancel
and the radiative corrections are described by the parton structure functions
which in general can be presented by the expression
%fig3
\begin{figure}
\begin{center}
  \begin{picture}(60,60)
  \Photon(30,0)(30,15){3}{3}
  \ArrowLine(60,15)(30,15)
  \ArrowLine(30,15)(30,45)
  \ArrowLine(30,45)(60,45)
  \Gluon(0,60)(30,45){3}{5}
  \end{picture}
\hspace*{1cm}
%  \hfill
  \begin{picture}(60,60)
  \Photon(30,0)(30,15){3}{3}
  \ArrowLine(60,15)(30,15)
  \ArrowLine(30,15)(30,45)
  \ArrowLine(30,45)(60,45)
  \Gluon(45,45)(45,15){3}{5}
  \Gluon(0,60)(30,45){3}{5}
\end{picture}
\hspace*{1cm}
%  \hfill
  \begin{picture}(60,60)
  \Photon(30,0)(30,15){3}{3}
  \ArrowLine(60,15)(30,15)
  \ArrowLine(30,15)(30,45)
  \ArrowLine(30,45)(60,45)
  \Gluon(30,30)(60,30){3}{5}
  \Gluon(0,60)(30,45){3}{5}
\end{picture}
\hspace*{1cm}
%  \hfill
  \begin{picture}(60,60)
  \Photon(30,0)(30,15){3}{3}
  \ArrowLine(60,15)(30,15)
  \ArrowLine(30,15)(30,30)
  \ArrowLine(30,30)(60,30)
  \Gluon(30,45)(60,45){3}{5}
  \Gluon(30,45)(30,30){3}{3}
  \Gluon(10,60)(30,45){3}{5}
\end{picture}
\vspace*{5mm}
  \caption{Contributions to the proces $\gamma^* + g \rightarrow
           '{\rm X}'$ contributing to the partonic structure function
           $\hat g_{1,g}$.}
  \label{fig:3}
\end{center}
\end{figure}
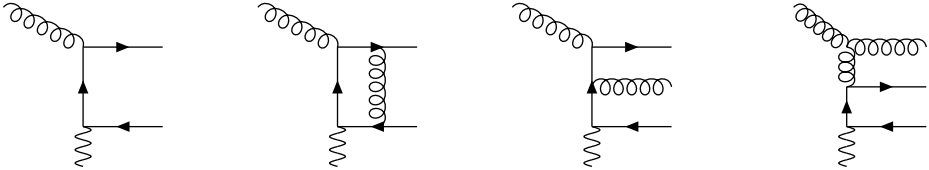
%--------------------------------------------
\begin{eqnarray}
\label{eqn:27}
&& \hat g_{1,i}^{r,(n)}(\frac{Q^2}{\mu^2},\frac{1}{\varepsilon})=\delta_{qi}
+ \hat a_s S_{\varepsilon} (\frac{Q^2}{\mu^2})^{\varepsilon/2}
\Big [- \frac{1}{\varepsilon} \Delta \gamma_{qi}^{(n),(0)} 
+  \Delta c_{1,i}^{(n),(1)} 
+ \varepsilon \Delta c_{1,i}^{\varepsilon,(n),(1)} \Big ]
\nonumber\\[2ex]
&& + {\hat a}_s^2 S_{\varepsilon}^2 (\frac{Q^2}{\mu^2})^{\varepsilon}
\Big [\frac{1}{\varepsilon^2} \Big \{ \frac{1}{2} \Delta \gamma_{qj}^{(n),(0)}  
\Delta \gamma_{ji}^{(n),(0)} + \beta_0 \Delta \gamma_{qi}^{(n),(0)} \Big \}
\nonumber\\[2ex]
&& + \frac{1}{\varepsilon} \Big \{- \frac{1}{2} \Delta \gamma_{qi}^{(n),(1)}
-2 \beta_0 \Delta c_{1,i}^{(n),(1)}
- \Delta c_{1,j}^{(n),(1)} \Delta \gamma_{ji}^{(n),(0)} \Big \}
\nonumber\\[2ex]
&& + \Delta  c_{1,i}^{(n),(2)} - 2 \beta_0 \Delta c_{1,i}^{\varepsilon,(n),(1)} 
- \Delta c_{1,j}^{\varepsilon,(n),(1)} \Delta \gamma_{ji}^{(n),(0)}
\Big ] \,,
\end{eqnarray}
with $r={\rm NS,S}$.
The above expression still contains ultraviolet and collinear divergences
both represented by the pole terms $(1/\varepsilon)^k$. The former are removed 
by coupling constant renormalization (see Eq. (\ref{eqn:17})). The 
residues of the collinear divergences are usually denoted by the DGLAP spin
splitting functions $\Delta P_{ij}$ which are related to the anomalous 
dimensions via a Mellin transform i.e.
\begin{eqnarray}
\label{eqn:28}
\Delta \gamma_{ij}^{(n)} = - \int_0^1\,dz\,z^{n-1}\, \Delta P_{ij}(z) \,.
\end{eqnarray}
The collinear divergences are removed by applying mass factorization which 
proceeds as follows
\begin{eqnarray}
\label{eqn:29}
&& \Delta {\bar {\cal C}}_{1,q}^{\rm NS,(n)} = (Z_{qq}^{5,{\rm NS},(n)})^{-1} 
({\bar \Gamma}^{-1})_{qq}^{{\rm NS},(n)} \hat g_{1,q}^{{\rm NS},(n)}
\nonumber\\[2ex]
&& \Delta {\bar {\cal C}}_{1,q}^{{\rm S},(n)} = (Z_{qq}^{5,{\rm S},(n)})^{-1} 
\Big [({\bar \Gamma}^{-1})_{qq}^{{\rm S},(n)} \hat g_{1,q}^{{\rm S},(n)} 
+ ({\bar \Gamma}^{-1})_{gq}^{{\rm S},(n)} \hat g_{1,g}^{{\rm S},(n)} \Big ]
\nonumber\\[2ex]
&& \Delta {\bar {\cal C}}_{1,g}^{{\rm S},(n)} =
({\bar \Gamma}^{-1})_{qg}^{{\rm S},(n)} \hat g_{1,q}^{{\rm S},(n)}  
+ ({\bar \Gamma}^{-1})_{gg}^{{\rm S},(n)} \hat g_{1,g}^{{\rm S},(n)} \,,
\end{eqnarray}
and the transition functions $({\bar \Gamma}^{-1})_{ij}^{r,(n)}$ are given by
(see Eq. (\ref{eqn:22}))
\begin{eqnarray}
\label{eqn:30}
({\bar \Gamma}^{-1})_{qq}^{{\rm NS},(n)}={\bar Z}_{qq}^{{\rm NS},(n)} 
\qquad , \qquad
({\bar \Gamma}^{-1})_{ij}^{{\rm S},(n)}={\bar Z}_{ij}^{{\rm S},(n)} \,.
\end{eqnarray}
Using these expressions the longitudinal spin coefficient functions
can be written as 
\begin{eqnarray}
\label{eqn:31}
&& \Delta {\bar {\cal C}}_{1,i}^{r,(n)}(\frac{Q^2}{\mu^2}) = \delta_{qi}  
+ a_s \Big [- \frac{1}{2} 
\Delta \gamma_{qi}^{(n),(0)}
\ln(\frac{Q^2}{\mu^2}) + \Delta {\bar c}_{1,i}^{(n),(1)} \Big ]
\nonumber\\[2ex]
&& + a_s^2 \Big [ \Big \{ \frac{1}{8} \Delta \gamma_{qj}^{(n),(0)} 
\Delta \gamma_{ji}^{(n),(0)}
+\frac{1}{4} \beta_0 \Delta \gamma_{qi}^{(n),(0)}\Big \}\ln^2(\frac{Q^2}{\mu^2})
\nonumber\\[2ex]
&& +\Big \{- \frac{1}{2} \Delta {\bar \gamma}_{qi}^{(n),(1)} - \beta_0 \Delta
{\bar c}_{1,i}^{(n),(1)} -\frac{1}{2} \Delta \gamma_{ji}^{(n),(0)} 
\Delta {\bar c}_{1,j}^{(n),(1)}\Big \}\ln(\frac{Q^2}{\mu^2})
+ \Delta {\bar c}_{1,i}^{(n),(2)} \Big ] \,,
\end{eqnarray}
so that they satisfy the renormalization group equations in
Eq. (\ref{eqn:12}). Notice that in the above expression the coefficients
$\Delta {\bar c}_{1,i}^{(n),(k)}$ differ from $\Delta c_{1,i}^{(n),(k)}$ 
in Eq. (\ref{eqn:27}). From the discussion above one infers that the
renormalization of the operator matrix elements determine the way one has
to perform the mass factorization on $\hat g_{1,i}^{r,(n)}$ and not vice versa.
The reason is that the renormalization of the former (but not of the latter) 
is ruled by the Ward-identities and some theorems which are violated
by the HVBM-prescription. This has forced us to introduce the additional 
constant $Z_{qq}^{5,r,(n)}$ in Eq. (\ref{eqn:20}) to restore the wanted 
properties on the level of the renormalized operator matrix elements 
presented in Eq. (\ref{eqn:23}). If we would have accidentally put 
$Z_{qq}^{5,r,(n)}=1$ the coefficient functions and the renormalized operator 
matrix elements get different anomalous dimensions and the relations in Eq.
(\ref{eqn:30}) would be violated. From Eqs. (\ref{eqn:24}), (\ref{eqn:25}) 
it also follows that the parton densities and the coefficient functions 
would have different anomalous dimensions. Hence the 
structure function $g_1(x,Q^2)$ (\ref{eqn:6}) would not satisfy 
Eq. (\ref{eqn:13}) anymore so that it is no longer a physical quantity.
Therefore this choice for $Z_{qq}^{5,r}$ is unacceptable.

\section{Fragmentation into polarized hadrons in electron-positron 
annihilation}
Single hadron (denoted by H) inclusive production in electron-positron
annihilation is given by the reaction
\begin{eqnarray}
\label{eqn:32}
e^-(l_1,\sigma_1) + e^+(l_2,\sigma_2) \rightarrow V(q) \rightarrow
{\rm H}(p,s) + '{\rm X}' \,.
\end{eqnarray}
Here we have introduced a similar notation to the one in reaction
(\ref{eqn:1}). However
the Bj{\o}rken scaling variable is defined by
\begin{eqnarray}
\label{eqn:33}
x= \frac{2p\cdot q}{Q^2}\,, \quad q^2=Q^2 > 0\,, \quad 0 < x \leq 1 \,,
\end{eqnarray}
for timelike momenta of the vector boson $V$.
The annihilation process can be depicted as in Fig. \ref{fig:1} where
now the incoming hadron is outgoing and the electron in the final state
becomes a positron in the initial state. In the case the incoming electron
in reaction (\ref{eqn:32}) is longitudinally polarized downwards, i.e.
$\sigma_1=\downarrow$, and the positron is unpolarized then one can simplify 
the cross section when the process becomes purely electromagnetic. In this 
case $V=\gamma$ and we get
\begin{eqnarray}
\label{eqn:34}
\frac{d\sigma^{{\rm H}(\downarrow)}(\downarrow)}{dx\,d\cos\theta}-
\frac{d\sigma^{{\rm H}(\uparrow)}(\downarrow)}{dx\,d\cos\theta}=
N_C\frac{\pi \alpha^2}{Q^2} \,\cos\, \theta \, g_1^{\rm H}(x,Q^2) \,.
\end{eqnarray}
The above expression represents the difference between the cross sections
where the detected hadron H is polarized parallel $s=\downarrow$ or 
anti-parallel $s=\uparrow$ with respect to the electron. Further $N_C$  
denotes the colour factor 
in $SU(N_C)$ and $\theta$ is the polar angle describing the direction of the
momentum of H in the C.M. frame with respect to the incoming electron. Notice
that the hadron fragmentation function $g_1^{\rm H}(x,Q^2)$ can be also 
measured in unpolarized electron-positron
scattering provided reaction (\ref{eqn:32}) is mediated by the Z-boson. Here
it appears via the axial-vector coupling of this boson to the lepton-pair.
The above hadron fragmentation function shows up in the anti-symmetric part
of the hadronic structure tensor
\begin{eqnarray}
\label{eqn:35}
W_{\mu\nu}(p,q,s) = \frac{1}{4\pi} \int\, d^4 z \,e^{iq\cdot z}
\langle 0 \mid J_{\mu}(z)\mid {\rm H}(p,s),{\rm X} \rangle  
\langle {\rm H}(p,s),{\rm X} \mid J_{\nu}(0) \mid 0 \rangle \,,
\end{eqnarray}
which can be decomposed in the same way as shown for Eq. (\ref{eqn:5}) in 
deep-inelastic scattering.
Since H is exclusive the above expression is not a
Fourier transform of a curent-current correlation function which implies that
we cannot insert the lightcone expansion of Eq. (\ref{eqn:15}). Therefore we 
can only work in the QCD improved parton model picture in which the 
fragmentation function has the following form
\begin{eqnarray}
\label{eqn:36}
&& g_1^{\rm H}(x,Q^2)=\frac{1}{n_f}\sum_{k=1}^{n_f} e_k^2 \int_x^1
\frac{dz}{z}\Big [
\Delta D_q^{\rm H,S}\Big (\frac{x}{z},\mu^2 \Big ) \Delta {\cal C}_{1,q}^{\rm S}
\Big (z,\frac{Q^2}{\mu^2} \Big )
\nonumber\\
&& + \Delta D_g^{\rm H,S}\Big (\frac{x}{z},\mu^2 \Big )
\Delta {\cal C}_{1,g}^{\rm S} \Big (z,\frac{Q^2}{\mu^2} \Big)
+ n_f \Delta D_k^{\rm H,NS}\Big (\frac{x}{z},\mu^2 \Big )
\Delta {\cal C}_{1,q}^{\rm NS}\Big (z,\frac{Q^2}{\mu^2} \Big )
\Big ] \,.
\end{eqnarray}
The spin parton fragmentation densities denoted by $\Delta D_i^{\rm H}(z,\mu^2)$
are the analogues of the parton densities in Eq. (\ref{eqn:6}) and
they satisfy the same renormalization group equations. However beyond lowest
order the anomalous dimensions are different for these two densities 
(see \cite{cfp}, \cite{fupe}, \cite{stvo}). The anomalous dimensions ruling 
the evolution of the spin fragmentation densities have recently been  
calculated up to second order in \cite{stvo}.
To obtain the timelike spin coefficient functions one has to calculate
the timelike photon analogues of the graphs in Figs. \ref{fig:2},
\ref{fig:3}. Here the incoming quark and gluon become outgoing and they now
fragment into the hadron H. Furthermore the 
spacelike photon turns into a timelike one. The calculation of these 
coefficient functions was recently done up to second order
in \cite{rijk}. It proceeds in the same way as for deep inelastic scattering
in section 1 where again the HVBM prescription for the $\gamma_5$-matrix 
is chosen. After having calculated the parton fragmentation functions denoted
by ${\hat g}_{1,i}^{{\rm H},r,(n)}$ one has to perform mass factorization 
analogously to Eq. (\ref{eqn:29}). However our calculation reveals that the 
renormalization
constant $Z_{qq}^{5,r}$ is different for timelike (fragmentation function)
and spacelike (structure function) quantities. In \cite{rijk} one has 
obtained the non-singlet part of this constant for the timelike (T)  
process (\ref{eqn:32}) by computing the ratio
\begin{eqnarray}
\label{eqn:37}
Z_{qq}^{5,{\rm NS,T},(n)} = \frac{ \hat g_{1,q}^{{\rm H,NS},(n)}
(Q^2/\mu^2,1/\varepsilon)}
{\hat {\cal F}_{3,q}^{{\rm H,NS},(n)}(Q^2/\mu^2,1/\varepsilon)} 
\mid_{\mu^2=Q^2} \,,
\end{eqnarray}
where $\hat {\cal F}_{3,q}^{{\rm H,NS},(n)}$ is the parton fragmentation 
function 
in unpolarized electron-positron annihilation. It arises due to the 
axial-vector coupling of the Z-boson to the outgoing quark anti-quark pair
(see Fig. \ref{fig:1}). If the $\gamma_5$ anticommutes with the other 
$\gamma$-matrices one can show that the coefficient
functions $\Delta {\cal C}_{1,q}^{\rm NS}$ and 
${\cal C}_{3,q}^{{\rm NS},(n)}$ are equal up to order $\alpha_s^2$. 
Since the HVBM prescription destroys
this property for $\hat g_{1,q}^{{\rm H,NS},(n)}$ we have to multiply the
latter by $Z_{qq}^{5,{\rm NS,T}}$ in order to obtain the correct coefficient 
functions. If we assign to 
the constant in Eq. (\ref{eqn:20}) the superscript S (here spacelike), the 
following relation holds for the inverse Mellin transforms 
\begin{eqnarray}
\label{eqn:38}
Z_{qq}^{5,{\rm NS,T}} (z) = -z Z_{qq}^{5,{\rm NS,S}}
(\frac{1}{z}) + a_s^2 \Big [\beta_0 z_{qq}^{{\rm NS},(1)}(z) \ln z\Big ] \,.
\end{eqnarray}
The above equalitity demonstrates the breakdown of the Gribov-Lipatov 
relation \cite{grli} in order $\alpha_s^2$. The above relation is also found 
for the timelike and spacelike non-singlet splitting functions 
$P_{qq}^{\rm NS}$ in unpolarized scattering in Eqs. (6.37)
and (6.38) of \cite{cfp} where $z_{qq}^{(1)}$ is replaced by the lowest order
DGLAP splitting function $P_{qq}^{{\rm NS},(0)}$. Notice that the same 
relation holds for the spin splitting functions because 
$\Delta P_{qq}^{\rm NS} = P_{qq}^{\rm NS}$ (see \cite{stvo}). The dependence
of $Z_{qq}^{5,{\rm NS}}$ on the quantity under consideration reveals
that it is not an universal constant. Therefore aside from the pole term 
already mentioned above Eq. (\ref{eqn:23}), it does not represent a genuine
renormalization constant in the usual sense.\\[5mm]
\noindent
Acknowledgements\\

We would like to thank J. Smith and Y. Matiounine for the careful reading of
the manuscript and for giving us some useful comments.


\begin{thebibliography}{99}
%Ref(1)
\bibitem{mene}
R. Mertig and W.L. van Neerven, Z. Phys. {\bf C70} (1996) 637.
%Ref(2)
\bibitem{vogel}
W. Vogelsang, Phys. Rev. {\bf D54} (1996) 2023; ibid. Nucl. Phys. {\bf B475}
(1996) 47.
%Ref(3)
\bibitem{msn}
Y. Matiounine, J. Smith and W.L. van Neerven, ITP-SB-02-98, INLO-PUB-02/98,
hep-ph/98 .
%Ref(4)
\bibitem{hove}
G. 't Hooft and M. Veltman, Nucl. Phys. {\bf B44} (1972) 189.
%Ref(5)
\bibitem{brma}
F. Breitenlohner and D. Maison, Commun. Mat. Phys. {\bf 52} (1977) 11, 39, 55.
%Ref(6)
\bibitem{akde}
D. Akyeampong and R. Delbourgo, Nuovo Cimento {\bf 17A} (1973) 578; {\bf 18A}
(1973) 94; {\bf 19A} (1974) 219.
%Ref(7)
\bibitem{larin}
S.A. Larin, Phys. Lett. {\bf B303} (1993) 113, {\bf B334} (1994) 192.
%Ref(8)
\bibitem{adba}
S.L. Adler and W. Bardeen, Phys. Rev. {\bf 182} (1969) 1517.
%Ref(9)
\bibitem{zine}
E.B. Zijlstra and W.L. van Neerven, Nucl. Phys. {\bf B417} (1994) 61; Erratum:
Nucl. Phys. {\bf B426} (1994) 245.
%Ref(10)
\bibitem{cfp}
G. Curci, W. Furmanski and R. Petronzio, Nucl. Phys. {\bf B175}
(1980) 27.
%Ref(11)
\bibitem{fupe}
W. Furmanski and R. Petronzio, Phys. Lett. {\bf B97} (1980) 437;
Z. Phys. {\bf C11} (1982) 293.
%Ref(12)
\bibitem{stvo}
M. Stratmann and W. Vogelsang, Nucl. Phys. {\bf B496} (1997) 41.
%Ref(13)
\bibitem{rijk}
P.J. Rijken and W.L. van Neerven, INLO-PUB-11/97, hep-ph/9711335.
%Ref(14)
\bibitem{grli}
V.N. Gribov and L.N. Lipatov, Sov. J. Nucl. Phys. {\bf 15} (1972) 438; 675.
\end{thebibliography}
\end{document}